 \documentclass[smallabstract,smallcaptions]{dccpaper}

\usepackage{epsfig}
\usepackage{amsmath}
\usepackage{amssymb}
\usepackage{enumitem}
\usepackage{color}
\usepackage{url}
\newlength{\figurewidth}
\newlength{\smallfigurewidth}
\usepackage{comment}
\usepackage{hyperref}

\begin{document}

\title
{\large
\textbf{Neural JPEG: End-to-End Image Compression Leveraging a Standard JPEG Encoder-Decoder}
}

\author{Ankur Mali\textsuperscript{$\star$},
Alexander G. Ororbia\textsuperscript{\textdagger}, Daniel Kifer \textsuperscript{$\star$},
C. Lee Giles\textsuperscript{$\star$}\\ [0.5em]
\textsuperscript{$\star$\ The Pennsylvania State University, University Park, PA, 16802, USA}\\
\textsuperscript{\textdagger Rochester Institute of Technology, Rochester, NY, 14623, USA}\\
}


\maketitle
\thispagestyle{empty}

\begin{abstract}
Recent advances in deep learning have led to superhuman performance across a variety of applications. Recently, these methods have been successfully employed to improve the rate-distortion performance in the task of image compression. 
However, current methods either use additional post-processing blocks on the decoder end to improve compression or propose an end-to-end compression scheme based on heuristics. For the majority of these, the trained deep neural networks (DNNs) are not compatible with standard encoders and would be difficult to deply on personal computers and cellphones. 
In light of this, we propose a system that learns to improve the encoding performance by enhancing its internal neural representations on both the encoder and decoder ends, an approach we call Neural JPEG. 
We propose frequency domain pre-editing and post-editing methods to optimize the distribution of the DCT coefficients at both encoder and decoder ends in order to improve the standard compression (JPEG) method. Moreover, we design and integrate a scheme for jointly learning quantization tables within this hybrid neural compression framework.Experiments demonstrate that our approach successfully improves the rate-distortion performance over JPEG across various quality metrics, such as PSNR and MS-SSIM, and generate visually appealing images with better color retention quality. 
\end{abstract}

\Section{Introduction}
\label{sec:intro}

Over the years there has been a rapidly increasing use of the world wide web, propelled by the ever growing popularity of social media, to transmit visual signals, e.g., pictures and video, of ever rising resolution and quality.
As a result, there is an equally increasing need to improve our ability to compress visual data.

Recently, in response to this problem, research in deep neural networks (DNNs) has begun to turn its attention to improving the rate-distortion performance of image compression frameworks.
Current efforts have presented impressive results with end-to-end image compression systems using DNNs in a variety of ways \cite{rippel2017real,cheng2019learning,yang2020heirarchical}. 
Although powerful, these systems require carefully designing and training effective encoding/decoding functions as well as quantizers. Other methods perform post-editing and apply DNNs in an attempt to reduce the compression artifacts on the decoder end \cite{wang2017novel,wang2018multi}. Though this set of approaches has yielded impressive results, they unfortunately require a specifically trained decoder during the post-processing stage or a complex DNN-based decoder. As such they are not supported by the commonly used image viewers in most computers and smartphones. 
In addition, there is no guarantee that the compression results uncovered would hold when the input data distribution shifts, e.g., images of a completely different kind are presented to the system, presenting a challenge to DNN-driven approaches.
Recent approaches that craft hybrid decoders \cite{ororbia2019dcc,mali2020dcc,dcc2021} have presented a very promising direction, yet they struggle to operate well at the lowest bit rate, given that the quantized signals they must work with to reconstruct the original input signal are extremely sparse. 
Another promising alternative is to design a hybrid encoder that enhances encoder signals resulting in better compression even at the lowest bit rates \cite{liu2016cu,strumpler2020learning}.  However, these methods fail to remove artifacts at the decoder end, thus compromising compression quality in various situations. 

In the spirit of the recent efforts to construct DNN-driven compression system, this paper proposes a simple approach to overcome the above mentioned shortcomings by handcrafting an image encoder-decoder system that exploits the inherent sparsity in quantization space while plugging into an existing image method.
Specifically, we focus on the most common and widely-used standard image compression method, JPEG, though our framework is general enough that it could be leveraged/extended to work with others, e.g., JP2 or PGF.
In this study, our approach improves the rate-distortion performance of JPEG by enhancing the latent representations acquired by its neural encoder and decoder by efficiently learning the DCT coefficients while at the same time ensuring that the bitstreams are decodable even with a standard decoder. 
Specifically, we construct a system that leverages an encoder and decoder that are each driven by sparse recurrent neural networks (SMRNNs) trained within the effective framework of neural iterative refinement \cite{ororbia2019dcc} -- the recurrent encoder learns to ``pre-edit'' an input image in the frequency domain (producing values that serve as the necessary DCT coefficients) while the recurrent decoder learns to reduce artifacts in the reconstructed image.
To further boost our rate-distortion performance at the lowest bitrate, we overcome the limitations of handcrafted codecs built into JPEG and JPEG-2000, which rely on quantization driven by fixed transformation matrices and entropy encoding, 
by developing a scheme that jointly learns the quantization table with the recurrent encoder.
As a result, since our approach learns the quantization table, the entire hbyrid system could be viewed as a differentiable JPEG pipeline.

In summary, our contributions are as follows:
\begin{itemize}[noitemsep,nolistsep]
    \item We extend on prior work and improve system rate-distortion performance by optimizing the JPEG encoder in the frequency domain.
    \item We facilitate better coefficient construction at the decoder end by optimizing the JPEG decoder. 
    \item A sparse recurrent network (Neural JPEG) is adapted to learn how to edit the DCT coefficients at both decoder and encoder ends. 
    \item A learnable quantization table that is optimized jointly with the sparse recurrent encoder/decoder to improve rate-distortion performance, yielding an end-to-end, differentiable JPEG compression system.
\end{itemize}

\Section{Related Work}
\label{sec:related_work}

Widely-used lossy image compression methods such as JPEG and JPEG-2000 (JP2) employ a combination of fixed transformations using entropy-based encodings to achieve better compression \cite{takamura1994coding}.  This is suitable for real-time processing when memory and computational efficiency are needed.  Recently,  DNN approaches have been shown to outperform these traditional methods in the task of image compression \cite{balle2017end, Toderici2016Variable}. However, most of this work has focused on designing end-to-end systems that reconstruct images in a two-dimensional space using architectural building-block models \cite{oord2016pixel} such as auto-encoders , convolutional networks, and recurrent networks. 
Some early work crafted a framework based on variational autoencoders \cite{gregor2016conceptualcompression} that results in improve rate-distortion. Other complementary work \cite{toderici2015,toderici2016full,johnston2018improved} , which outperformed classical techniques  (at low bit rates) without harming perceptual quality, set the widely-adopted practice for using deep artificial neural networks (ANNs) in compression. 
Later methods that followed focused on using convolutional networks or generative adversarial networks (GANs) \cite{Theis2017,rippel2017real,balle2018variational,agustsson2017soft}.  Recent work has used spatial-temporal energy compaction \cite{cheng2019learning}, other energy compaction-based techniques \cite{chen19}, and filter-bank-based convolution networks \cite{dcc19}.   
Most of these end-to-end solutions have been designed to extract better latent representations and/or eliminate redundancies in the compression process. 
A more straightforward compression approach considered redundancy at the decoder side of the system and attempted to decompress by designing an iterative hybrid recurrent decoder \cite{ororbia2019dcc,mali2020dcc, dcc2021}.  Similarly, a standard encoder can be replaced with another DNN to enhance the model's internal neural representations and decode information while only using a standard decoder both in the pixel \cite{talebi2021better} and frequency domains \cite{strumpler2020learning}.

\Section{Neural JPEG}
\label{sec:neural_jpeg}

\subsection{The JPEG Algorithm}
We start by briefly introducing the workflow of the JPEG algorithm.
The first step employs conversation of the input image from RGB to the YCbCr colorspace. Next, the image $\mathbf{I}$ (of size $N \times M$ pixels) is divided into $N$ non-overlapping blocks of size $N_i \times M_i$ pixels. The discrete cosine transform (DCT) is then applied to convert each block into the frequency domain. We denote the DCT coefficients of any given block $(n,m)$, where $n \in 0...(N/N_i)$ and $m \in 0...(M/M_i)$, 
for the luminance channel $Y$ with $\mathbf{F}^{(Y)} = \mathbf{I}[n,m] \in \mathbb{R}^{8 \times 8}$ and accordingly for chrominance channels $Cb, Cr$. 
Furthermore, the DCT or DWT (in JPEG 2000) coefficients are quantized using two quantization tables: $\mathbf{Q}^{(L)} \in \mathbb{R}^{8 \times 8} $ for the luminance channel $Y$ and $\mathbf{Q}^{(C)} \in \mathbb{R}^{8 \times 8}$ for the chrominance channels $Cb,Cr$ followed by a rounding function:
\begin{align}
    \Hat{ \mathbf{Z} }_{u,v}^{(Y)} = \Bigg \lfloor \dfrac{ \mathbf{F}_{u,v}^{(Y)}} {Q_{u,v}^{(L)}} \Bigg \rceil , \;
    \Hat{ \mathbf{Z} }_{u,v}^{(Cb)} = \Bigg \lfloor \dfrac{ \mathbf{F}_{u,v}^{(Cb)}} {Q_{u,v}^{(C)}} \Bigg \rceil , \;
    \Hat{ \mathbf{Z} }_{u,v}^{(Cr)} = \Bigg \lfloor \dfrac{ \mathbf{F}_{u,v}^{(Cr)}} {Q_{u,v}^{(C)}} \Bigg \rceil , \text{ for } u,v \in [1,8] \mbox{.} \label{eqn:jpeg_dct} 
\end{align}
Finally, these quantized DCT coefficients are passed to an entropy coding module to finish compressing the input image.

\begin{figure}[!t]
\centering
\includegraphics[width=0.99\linewidth]{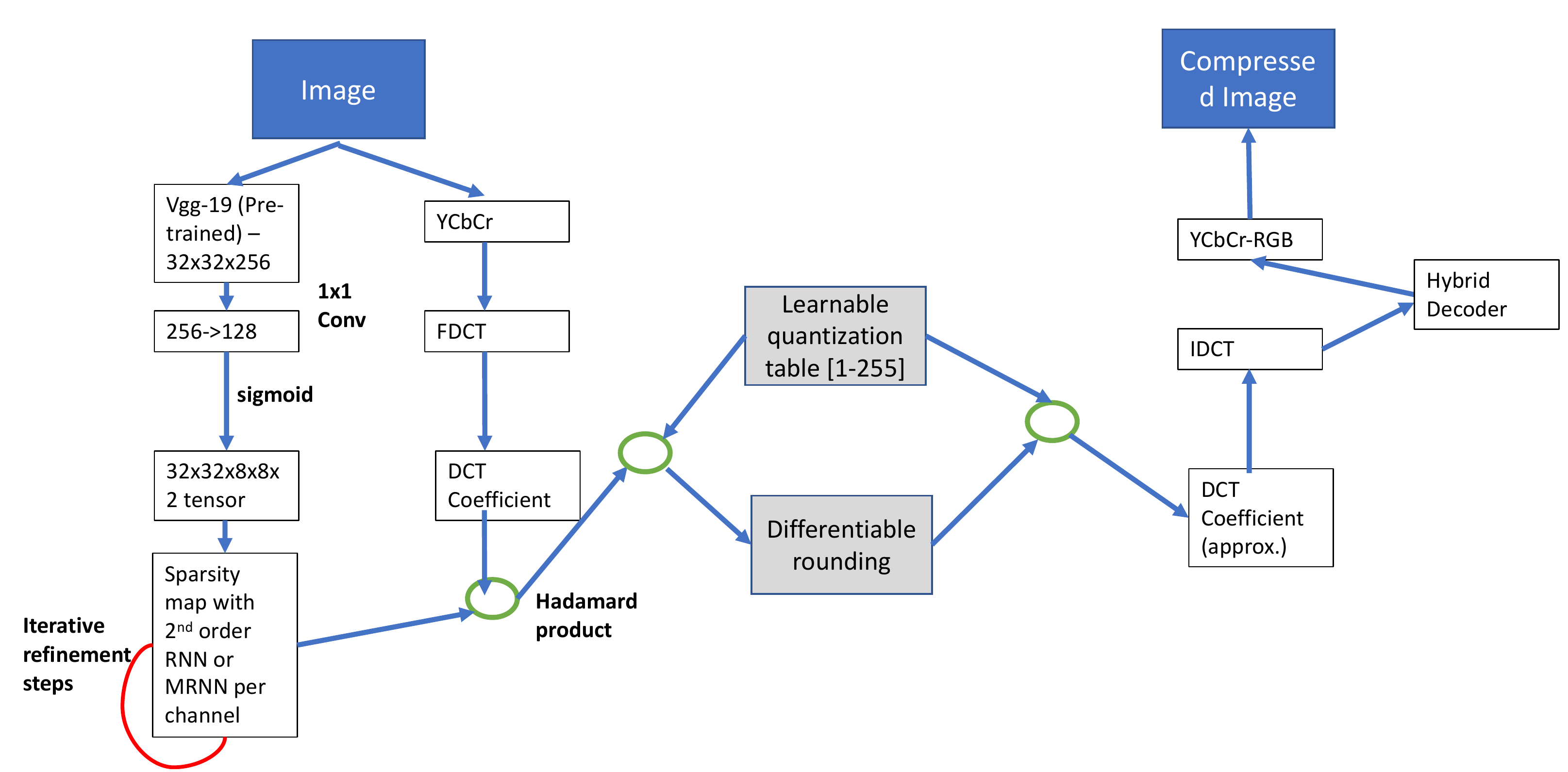}
\caption{The end-to-end Neural JPEG architecture for image compression.}
\label{fig:net}
\vspace{-0.7cm}
\end{figure}

\subsection{The Neural JPEG Architecture} 
\label{sec:neural_jpeg}

Our complete Neural JPEG architecture, which integrates elements from \cite{Shin2017JPEGresistantAI} and \cite{ororbia2019dcc}, leverages a recurrent encoder model $E_\Theta(\mathbf{I})$ that drives/interacts with the encoder of standard JPEG, a differentiable quantization table and rounding module, and a recurrent decoder model $D_\Theta(\Hat{\mathbf{Z}})$ that drives/interacts with the decoder of standard JPEG. All of these components/modules are optimized jointly yielding a one-time training end-to-end differentiable JPEG pipeline.
A graphical representation of the proposed Neural JPEG is depicted in Figure \ref{fig:net}.
Formally, the Neural JPEG architecture can be summarized as the following functions:
\begin{equation}
     \Hat{\mathbf{Z}} = E_\theta( \mathbf{I} ), \quad \hat{\mathbf{I}} = D_\theta(\Hat{ \mathbf{Z} }) \implies \quad \hat{ \mathbf{I} } = D_\theta(E_\theta(\mathbf{I}) ).
     \label{autoenc}
\end{equation}
where $\hat{\mathbf{I}}$ and $\Hat{\mathbf{Z}}$ represents the reconstructed image and quantized DCT coefficients respectively.

\paragraph{Editing with Sparse RNNs:}
Inspired by \cite{talebi2020better} and \cite{strumpler2020learning}, we design an approach that uses a neural models to either pre-edit (or iteratively process) an image $\mathbf{I}$ before the quantization step of JPEG or post-edit the inverse DCT coefficients before converting back to the reconstructed image $\Hat{\mathbf{I}}$.
Specifically, for pre-editing, our neural encoder $E_\Theta(\mathbf{I})$ proceeds according to the following steps to produce a set of ``edit'' weights:
\begin{enumerate}[noitemsep,nolistsep]
    \item Run $\mathbf{I}$ through vision model $\mathbf{H}' = \text{DS}( f_V(\mathbf{I};\Theta_V)$ ), where $\text{DS}()$ is the down-sampling operator. Note that $f_V$ in this paper was chosen to be a pre-trained VGGNet-19 \cite{simonyan2014very} and $\Theta_V$ are its parameters. A $1 \times 1 $ convolution is then applied to $\mathbf{H}'$ to reduce its channel dimension to $128$ and the final output is $\mathbf{H} = \sigma(\mathbf{H}')$ where $\sigma$ is the sigmoid activation function to limit the output values to the range $[0,1]$.
    The $128$ output channels of $\mathbf{H}$ are finally reshaped to a $8 \times 8 \times 2$ tensor for each block which is then split into two $8 \times 8$ matrices (one for luminanace and one for chrominance), i.e., $\mathbf{H}_L$ and $\mathbf{H}_C$. 
    \item Use the sparse multiplicative RNN (SM-RNN) component to process both $\mathbf{H}_L$ and $\mathbf{H}_C$ (in parallel) for $K$ steps (which is the process of iterative refinement, adapted from \cite{ororbia2019dcc}, using the multiplicative RNN model proposed in \cite{sutskever2011mrnn}). This SM-RNN produces the desired  activation map according to the  dynamics:
    \begin{align}
        \mathbf{f}^L_k &= \text{diag}(\mathbf{W}_f \mathbf{H}_L) \cdot (\mathbf{V}_f \mathbf{z}^L_{k-1}) \\ \mathbf{z}^L_k &= \phi(\mathbf{V}_z \mathbf{f}^L_k + \mathbf{W}_z \mathbf{H}_L), \text{where } k = 0,1,...,K \label{eqn:mrnn}
    \end{align}
    where the above equation is simply repeated for $\mathbf{H}_C$ (just replace $L$ in Equation \ref{eqn:mrnn} with $C$) and $K = 3$ in this paper.
    The final set of sparse ``edit'' values are produced via the following:
    \begin{align}
        \mathbf{c}_L = \text{kWTA}( \mathbf{U} \mathbf{z}^L_K ), \; \mathbf{c}_C = \text{kWTA}( \mathbf{U} \mathbf{z}^C_K )
    \end{align}
    where $\text{kWTA}()$ is the $k$ winners-take-all function (setting all values that are less than the $k$ greatest neurons are set to zero).
\end{enumerate}
Note that the above is repeated for each of the $N \cdot M$ blocks. 
Before applying our quantization table, we multiply each DCT coefficient produced by the original JPEG encoder by its corresponding edit score obtained from the above process.


For the decoder end of our system, we utilize a SM-RNN similar to the encoder process described above but tie its weights to those of the SM-RNN encoder module (both models would have the same dimensionalities for their parameters given they operate in the same internal latent space). In short, after applying the inverse DCT transform to the outputs of the quantization table and rounding modules (which produce $\Hat{\mathbf{Z}}$), the SM-RNN decoder takes in the $\Hat{\mathbf{Z}}$, processes it $K$ times (in a process similar to the one depicted above) and finally produces the reconstructed image $\Hat{\mathbf{I}}$.


\paragraph{A Learnable Quantization Table and Rounding Module: }
In line with prior work \cite{strumpler2020learning} we also use the differentiable JPEG pipeline to learn the quantization tables. We replace attention mechanism with sparse RNN to better capture the importance of each representation associated with each channel.
We use $ Q_\theta^{(L)}$ and $  Q_\theta^{(C)}$ as optimization variables for luminance and chrominance. The range of quantization table values are clipped to $[1,255]$ which helps in optimization process.
Our overall approach in this stage is as follows:
\begin{equation}
\begin{gathered}
\Hat{ \mathbf{Z}}^{(Y)}[n,m] = \Bigg \lfloor  \mathbf{F}^{(Y)}[n,m]  \odot  SMRNN^{(L)}[n, m]  \odot \ {\Bar{ \mathbf{Q}}_\theta^{(L)}}\Bigg \rceil_{approx}\\
\Hat{ \mathbf{Z}}^{(Cr)}[n,m] = \Bigg \lfloor  \mathbf{F}^{(Cr)}[n,m]  \odot  SMRNN^{(C)}[n, m] \odot \ {\Bar{ \mathbf{Q}}_\theta^{(C)}}\Bigg \rceil_{approx}\\
  \Hat{ \mathbf{Z}}^{(Cb)}[n,m] = \Bigg \lfloor  \mathbf{F}^{(Cb)}[n,m]  \odot  SMRNN^{(C)}[n, m]  \odot \ {\Bar{ \mathbf{Q}}_\theta^{(C)}}\Bigg \rceil_{approx}\\
   \quad \text{ for } n \in [1,N], m \in [1,M],\\
   \text{with }
 \Bar{Q}_{u,v}^{(L)} = \dfrac{1} {Q_{u,v}^{(L)}}, \quad
  \Bar{Q}_{u,v}^{(C)} = \dfrac{1} {Q_{u,v}^{(C)}},
  \quad \text{ for } u,v \in [1,8].
\end{gathered}
\label{quantize}
\end{equation}
Here $\odot$ represents hadamard product. This The modification introduced above loses some features and is not recoverable at decoder end. Multiplying the DCT coefficients by a number $\leq$ 1 acts like a frequency filter, suppressing the higher frequency to get low-pass filter. By combining sparse weights with the DCT-coefficient we get a smoothing filter that is spatially adaptive and also applicable across various frequencies. 

For the rounding module, we remove the entropy encoding used in JPEG and replace the hard rounding operation with a differentiable 3rd order approximation:
\begin{equation}
\begin{gathered}
\lfloor \Hat{ \mathbf{Z} }  \rceil_{approx} =   \lfloor \Hat{ \mathbf{Z} }  \rceil + (\lfloor \Hat{ \mathbf{Z} }  \rceil -\Hat{ \mathbf{Z} } )^3.
\end{gathered}
\end{equation}


\subsection{Loss Formulation}
For designing end-to-end framework one needs to find a efficiently optimize for rate-distortion tradeoff. Additionally we introduced alignment loss that is responsible for controlling hybrid decoder.
For any given input image $ x$ and the reconstructed image $\hat { x}$, and learned parameters $\theta$, our loss function has the general form as follows:
\begin{equation}
 \begin{gathered}
 \mathcal{L}( x, \hat { x}; \theta) = \lambda \cdot \text{d}( x, \hat { x}) + (1-\lambda - 0.01) \text{r}( x, \hat { x}; \theta) + 0.01\text{al}( x, \hat { x}),\\
 \text{with }   x, \hat { x} \in \{ t \in \mathbb{R} \mid 0 \leq t \leq 255\}^{8N \times 8M  \times 3}, \quad \lambda \in \mathbb{R}^+,
 \end{gathered}
 \end{equation}
 where $\text{al}( x, \hat { x}; \theta)$ is alignment loss, $\text{r}( x, \hat { x}; \theta)$ is rate loss and $\text{d}( x, \hat { x})$ is distortion loss.
\noindent The parameter $\lambda$ determines the ratio of the triplet loss and hence balances alignment, distortion and rate. The ideal value for $\lambda$ is obtain based on validation performance.
 
\subsubsection{The Distortion Loss}
The distortion loss is responsible for measuring similarity between compressed and original images. To achieve this we use the combination MSE 
and LPIPS as follows:
\begin{equation}
\text{d}( x, \hat { x}) = \text{MSE}( x, \hat { x}) + \gamma \cdot \text{LPIPS}( x, \hat { x})
\end{equation}
 where we introduce $\gamma$ as the LPIPS modulating factor.
 
\subsubsection{The Rate Loss}
 We use the rate loss formulation proposed by \cite{strumpler2020learning} by replace attention map with sparse map obtained from SMRNN that is represented as follows:
\begin{equation}
 \text{r}( x; \theta) = \alpha(\Vert\bar{ Q}_\theta^{(L)}\Vert_1 + \Vert\bar{ Q}_\theta^{(C)}\Vert_1) + \beta (\text{mean}( SMRNN_\theta^{(L)}( x)) + \text{mean}( SMRNN_\theta^{(C)}( x)))
 \label{rate_loss}
  \end{equation}
with the mean function: $\text{mean}(  SMRNN) = \dfrac{1}{\vert \mathcal{P} \vert}\sum_{\vec p \in \mathcal{P}} SMRNN_{\vec p}$,
where $\mathcal{P}$ is the index set over all entries in the tensor $ SMRNN$.

\subsubsection{Alignment Loss}
The aligment loss is ensuring signals are robust at decoder end.
To achieve this we propose using combination of MSE and Mean Absolute Error (MAE) as follows:
\begin{equation}
 \text{al}( x, \hat { x}) = (1- \sigma) \text{MSE}( x, \hat { x}) + \sigma \cdot \text{MAE}( x, \hat { x})
  \end{equation}
where $\sigma=[0.1-0.4]$ (values chosen based on validation set).
Based on the triplet loss definitions above we can pose the optimization objective as: $\underset{\theta}{\min} \,\, \mathcal{L}( x, D_\theta(E_\theta(  x) ); \theta)$.

\Section{Experiments}

\subsection{Evaluation Metrics}

\noindent We use widely used compression optimization metric for measuring image similarity the Mean Squared Error (MSE) and the Peak Signal to Noise Ratio (PSNR). Similarly to \cite{Cavigelli_2017, strumpler2020learning} we define the MSE and PSNR for the tensors $ x, \hat{  x} \in hcal{X}$ of arbitrary dimension as follows ($x$ is also the input image $\mathbf{I}$):
\begin{equation}
\begin{gathered}
    \text{MSE}(  x, \hat{  x}) = \dfrac{1}{\vert \mathcal{P} \vert}\sum_{\vec p \in \mathcal{P}}( x_{\vec p} - \hat{ x}_{\vec p})^2\\
    \text{PSNR}( x, \hat{  x}) = 10 \log_{10} \left( \dfrac{255^2}{\text{MSE}(  x, \hat{  x})} \right)\\
\end{gathered}
\label{MSE}
\end{equation}
where  $\mathcal{P}$ is the set of pixel indices and  $x_{\vec p}, \hat{x}_{\vec p} \in [0,255]$, $\quad \forall$ $\vec p \in \mathcal{P}$.
To better measure the visual appeal of images we also use the Multi-Scale Structural Similarity (MS-SSIM) \cite{mssim}, converted to a logarithmic scale as follows:
\begin{equation}
    \text{MS-SSIM [dB]} = -10\log_{10}(1 - \text{MS-SSIM})
\end{equation}
Additionally, we also use DNN-friendly the Learned Perceptual Image Patch Similarity (LPIPS) \cite{zhang2018unreasonable} objective function.

\subsection{Datasets and Training Procedure}
The Neural JPEG network is trained on the dataset provided in prior work \cite{hasinoff, strumpler2020learning}. It consists of 3640 HDR images. For training, we use the merged HDR images and extract image patches of size $256$ obtained from random cropping. We follow the same extraction process and experimental protocol proposed by \cite{strumpler2020learning}
We evaluate our model on the Kodak dataset, consisting of 24 uncompressed images of size $768 \times 512$. Additionally, we validate our model on validation set from DIV2K \cite{div2k,Timofte_2017_CVPR_Workshops} containing 100 high quality images with 2040 pixels.
Our model is optimized using Adam with initial learning rate $1.0$ reduced to $10^{-8}$ using polynomial decay. We use batch size of $32$ for all experiments and performed grid search to find optimal hidden sizes for SMRNN, sparsity level $k$, and $\lambda$.
We use pre-trained VGG-19 model (trained on ImageNet) and fine-tune these layers while training. The  $1 \times 1$ convolutional layer is initialized using orthogonal matrices. We follow prior experimental protocol \cite{strumpler2020learning}, hence the quantization table variables in this work are also initialized uniformly in the interval $[1s,2s]$ and are limited to be in the range $[1s, 255s]$. Where the scaling factor is always $s > 0$ and in this experiments is set to is a $s=10^{-5}$. Then we can get the final quantization tables by multiplying factor by $s^{-1}$. We use standard evaluation metrics such as PSNR, MSE, MS-SSIM to report our model performance.

\begin{figure}
    \centering
    \includegraphics[width=0.99\linewidth]{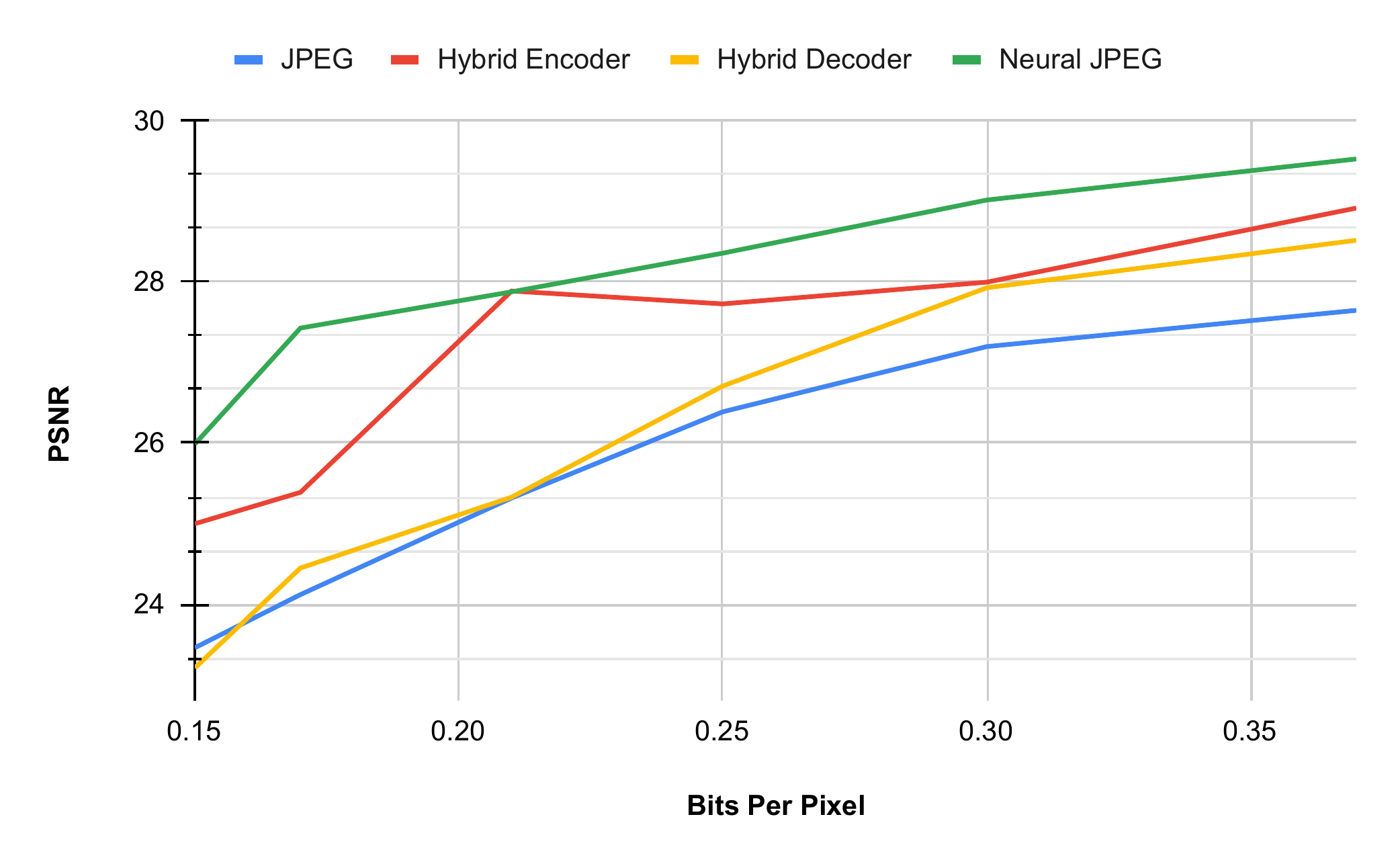}
    \caption{Performance of various compression model with various bit rate setting}
    \label{fig:bpp_result}
    \vspace{-0.5cm}
\end{figure}

\Section{Result and Discussion}
We evaluated our model on $2$ out of $6$ benchmarks used in prior work \cite{ororbia2019dcc} using $3$ metrics \cite{ma2016group}. These metrics are Peak Signal to Noise Ratio (PSNR), structural similarity (SSIM), and multi-scale structural similarity (MS-SSIM \cite{wang2004imagequality}, or $MS^3IM$. We compare our model against wide variety of compression approaches such pure neural-based GOOG\cite{toderici2016full} and E2E \cite{balle2017end}. We also compare to the models of  \cite{ororbia2019dcc} and \cite{strumpler2020learning}. Results are reported in Table \ref{results:benchmarks} -- we see all models perform stably, however, when the bit rates are reduced  (see Table \ref{results:benchmarks1}) all hybrid models start struggling, whereas neural-based models stay consistent. This suggests that the performance of these hybrid models is limited due to the fixed map used by JPEG. 
This support our hypothesis that at lower bit rates enhancing JPEG signals further improves performance and that Neural JPEG does this efficiently. 
Our model does struggles due to some artifacts, but these can be removed with JPEG artifact removal and we note that our overall model ensures the structure/content of image is kept intact. 
In other words, whenever we operate at the lowest bit rates (see Figure \ref{fig:bpp_result}), JPEG completely loses image information while the Neural JPEG makes up the difference since it can recover majority of missing chunks of features. 
Since JPEG quality based on PSNR at 0.25bpp is equivalent to Neural JPEG image quality at 0.19bpp, we conclude that even at the lowest bit rates, our model tries to remember majority of the signal.

\begin{table}[!t]\renewcommand{\arraystretch}{1.0}
\begin{center}
\caption{Test results for Kodak (bpp $0.38$), 8-bit Compression Benchmark (CB, bpp, $0.371$)}
\label{results:benchmarks1}

\resizebox{\textwidth}{!}{%
\begin{tabular}{|l||c|c|c|c||c|c|c|c|}
\hline
  & \multicolumn{3}{c||}{\textbf{Kodak}} & \multicolumn{3}{c|}{\textbf{CB 8-Bit}}\\
  \textbf{Model} & \textbf{PSNR} &  \textbf{SSIM} &  $MS^3IM$ & \textbf{PSNR} & \textbf{SSIM} & $MS^3IM$\\
  \hline
  \emph{JPEG} & 31.2190 & 0.7412 & 0.9011 & 32.7893 &  0.7921 & 0.9009 \\
  \emph{GOOG}-JPEG & 30.9821  & 0.7415 & 0.9016 & 31.899 &  0.7967 & 0.9012 \\
  \emph{E2E} (\textbf{Neural}) & 30.3471  & 0.7521 & 0.9021 & 31.769  & 0.8001 & 0.9016 \\
  \emph{MLP}-JPEG & 27.8325  & 0.8399 & 0.9444 & 27.8089 &  0.8371 & 0.9475 \\ 
 $\Delta$-\emph{RNN}-JPEG & 28.5093  & 0.8411 & 0.9487 & 28.0461  & 0.8403 & 0.9535 \\
 \emph{GRU}-JPEG & 28.5081  & 0.8400 & 0.9474 & 28.0446 &  0.8379 & 0.9533 \\
 \emph{Hybrid Decoder} - JPEG & 31.1282  & 0.7413 & 0.9011 & 32.7100 & 0.7920 & 0.9010 \\
 \emph{Hybrid Encoder} & 31.512  & 0.7520 & 0.9021 & 31.6712 & 0.8002 & 0.9016 \\
 \emph{Neural JPEG} (\textbf{Ours}) & 31.732  & 0.7521 & 0.9022 & 32.400 & 0.7926 & 0.9012 \\
 \hline
\end{tabular}
}
\end{center}
\end{table}

\begin{table}[!t]\renewcommand{\arraystretch}{1.0}
\begin{center}
\vspace{-0.55cm}
\caption{Test results for Kodak (bpp $0.37$), 8-bit Compression Benchmark (CB, bpp, $0.341$)}
\label{results:benchmarks}
\resizebox{\textwidth}{!}{%
\begin{tabular}{|l||c|c|c|c||c|c|c|c|}
\hline
  & \multicolumn{3}{c||}{\textbf{Kodak}} & \multicolumn{3}{c|}{\textbf{CB 8-Bit}}\\
  \textbf{Model} & \textbf{PSNR} &  \textbf{SSIM} &  $MS^3IM$ & \textbf{PSNR} & \textbf{SSIM} & $MS^3IM$\\
  \hline
  \emph{JPEG} & 27.6540 & 0.7733 & 0.9291 & 27.5481 &  0.8330 & 0.9383 \\
  \emph{GOOG}-JPEG & 27.9613  & 0.8017 & 0.9557 & 27.8458 &  0.8396 & 0.9562 \\
  \emph{E2E} (\textbf{Neural}) & 28.9420  & 0.8502 & 0.9600 & 28.0999  & 0.8396 & 0.9562 \\
  \emph{MLP}-JPEG & 27.8325  & 0.8399 & 0.9444 & 27.8089 &  0.8371 & 0.9475 \\ 
 $\Delta$-\emph{RNN}-JPEG & 28.5093  & 0.8411 & 0.9487 & 28.0461  & 0.8403 & 0.9535 \\
 \emph{GRU}-JPEG & 28.5081  & 0.8400 & 0.9474 & 28.0446 &  0.8379 & 0.9533 \\
 \emph{Hybrid Decoder} - JPEG & 28.5247  & 0.8409 & 0.9486 & 28.0461 & 0.8371 & 0.9532 \\
 \emph{Hybrid Encoder} & 28.8920  & 0.8411 & 0.9488 & 27.9211 & 0.8374 & 0.9534 \\
 \emph{Neural JPEG} (\textbf{Ours}) & 29.5247  & 0.8413 & 0.9489 & 27.8009 & 0.8375 & 0.9535 \\
 \hline
\end{tabular}
}
\end{center}
\vspace{-0.75cm}
\end{table}

\Section{Conclusions and Future Work}
Our experiments show that our approach, Neural JPEG, improves JPEG encoding and decoding through sparse RNN smoothing and learned quantization tables that are trained end-to-end in an differentiable framework. The proposed model leads to better compression/reconstruction at lowest bit rates when evluated using metrics such as MSE, PSNR and also using perceptual metrics (LPIPS, MS-SSIM) that are known to be much closer to human perception. Most importantly, the improved encoder-decoder remains entirely compatible with any standard JPEG algorithm but produces significantly better colors than standard JPEG. We have shown that we can achieve improvement without directly estimating the entropy of the DCT coefficients, only regularizing the sparse maps and quantization tables. 
In the future,  we wish to design an improved decoder that learns quantized signals from each color channel and uses a distribution-specific quantization table instead of a single differentiable quantization table.

\Section{References}
\bibliographystyle{IEEEbib}
\bibliography{zmain}

\end{document}